**Modelling resonant transmission in a Fabry-Perot/Surface-Plasmon microcavity**

Yago Arosa, Alejandro Doval, Raúl de la Fuente

Grupo de Nanomateriais, Fotónica e Materia Branda, Departamento de Física Aplicada, Universidade de Santiago de Compostela, E-15782, Santiago de Compostela, Spain

**Abstract**. Coupled surface plasmons arise in the surfaces of a dielectric layer between two metallic media when a dim wave propagating in the dielectric generates resonant free charge oscillations at the interfaces. Here, we consider surface plasmon resonance in a Fabry-Perot type cavity with plane metallic mirrors and an inner dielectric medium, optically less dense than the outer surrounding dielectric medium. The experimentally observed transmission as a function of both the angle of incidence of light and the wavelength is well modelled by an elementary transmittance function from which resonance conditions are obtained both in the Fabry-Perot and Surface-Plasmon regime.

**1. Introduction**

The aim of this work is to study a simple optical microcavity (MC) consisting of two flat metallic mirrors with a very thin dielectric medium between them. These three layers make up a conventional metal-dielectric-metal (MDM) structure. The atypical feature of the cavity under study is that the intracavity dielectric has a lower refractive index than that of the medium surrounding the cavity, which is a dielectric too. It is a typical Fabry-Perot (FP) cavity, but with two distinctive characteristics: the high absorption of its metallic mirrors, and its singular behaviour at high angles of incidence. This behaviour is due to the generation of coupled surface plasmons (CSP), collective charge oscillations at both metal-dielectric interfaces, by coupling external light to the cavity. To induce these plasmonic effects by means of attenuated total reflection (ATR) of a signal in the optical domain, the cavity length, or intracavity thickness, must be kept in the micron range.

The subject to be studied is outlined in Fig. 1, a cavity formed by an MDM structure between two high refractive index dielectrics. For simplicity, we consider a symmetrical configuration. From the inside, we have a dielectric layer of refractive index $n_L$ and thickness $d$ between two identical metallic films of thickness $d_M$ (typically $d_M < d$) deposited on a dielectric with refractive index $n_H > n_L$ (the subindices $L$ and $H$ stand for low and high, respectively). At low angles of incidence, the cavity exhibits the familiar optical characteristics of an etalon or Fabry-Perot interferometer with highly reflective mirrors (HRM) [1]. That is, a high light transmittance when specific resonance conditions are achieved, which paradoxically becomes more pronounced as the reflectivity of the mirrors increases. Furthermore, the shorter the cavity length, the more widely spaced these resonances are, both in the spectral and angular domain. As long as $n_H > n_L$, total reflection processes occur at the first cavity mirror if the incidence angle is large enough. If the second mirror is not sufficiently close (say $d$ of tens of microns), the light does not enter the cavity but is reflected with an evanescent tail in the internal medium. In the case of a transparent mirror (or no mirror), this reflection would be total (TIR, total internal reflection) [2, 3]. However, in the case of metallic mirrors, there is attenuated total reflection (ATR) caused by absorption in the metal, especially when the conditions of surface plasmon resonance (SPR) are verified [4]. Nevertheless, if the two mirrors are brought close enough, reflection can be frustrated, and an outstanding transmission can be obtained also under a resonance condition. Although these resonances are plasmonic in nature, the new condition is dependent on the intracavity thickness. It is associated to coupled charge oscillations at the inner surfaces of both metallic mirrors, called coupled surface plasmons (CSP) [5]. All this rich phenomenology associated with the MC is represented in figure 1b.

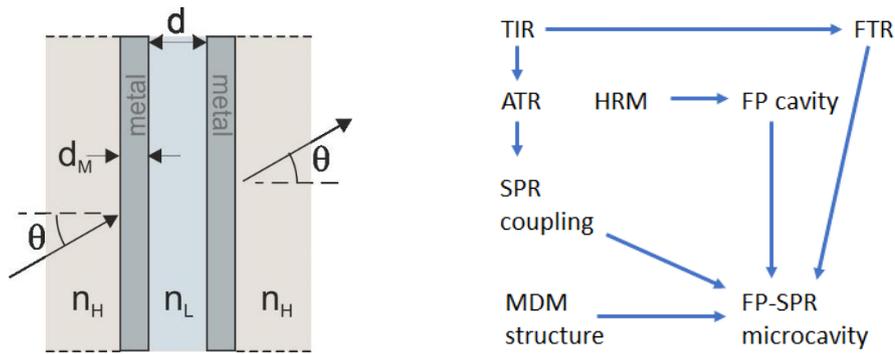

Figure 1. (a) Outline of a MC to be investigated; (b) Chart of phenomena involved in the MC. All acronyms are defined in the text.

Unlike most previous studies, which have focused on studying the reflectance of an MC and its plasmonic resonances, we consider here the properties of its transmission, which are fortunately simpler and perhaps even more surprising. For example, CSPs are associated with high transmission for intracavity thicknesses of several wavelengths, despite the high reflectance and absorptance of the mirrors. We carry out the study based on an analytical formula for transmittance that includes the surface plasmon (SP) regime and generalises the well-known formula applied in the FP regime. Typical resonance conditions, in line with the literature on plasmonics, are identified.

Beyond this kind of plasmonic structures, research on optical microcavities of various kinds and devoted to different applications has proliferated over the last decades [6, 7]. Focusing on MCs involving metallic layers, there were some pioneering numerical and experimental studies on the properties of their reflection [8, 9] and transmission [5]. These were more recently followed by some interesting (although not numerous) works on the characteristics of this kind of MCs [10, 11, 12, 13]. There are also studies on guiding in MDM plasmonic structures [14, 15, 16], from which one can deduce properties of MCs, even though they do not directly address reflection and transmission in cavities. The most studied applications are designing spectral filters [17, 18] and improving the resolution of SPR sensors [19, 20]. As a concrete example of the former, spectral resonances independent of the direction of the incident light have been achieved [21], and as an example of the latter, resonances have been used for the characterisation of nanometre-thick films [22].

## 2. Experimental results

The most direct way to insert light into an MC with high angles of incidence is by using a high refractive index coupling prism. Equivalently, since our intention is to study transmission properties, a similar prism can be used to collect the transmitted light. In our experimental setup, we use a supercontinuum light source (450 to 2400 nm spectrum) to illuminate the cavity, and a linear polariser to obtain a transverse magnetic (TM) beam, which is essential to excite surface plasmons. We used two right angle prisms made of HK9 glass with silver deposited on their largest faces using a high vacuum evaporator. The thickness of the metal films is approximately 45 nm, as measured by a profilometer. We place the prisms facing each other on their metallised faces, with a very small air gap between them. We mount this system on a rotating platform that allows us to change the angle of incidence. Our detection device is a homemade spectrometer, with a spectral range between 400 and 1100 nm. Additionally, a variable density filter is used to adjust the power of the light incident in the cavity, and a photodiode monitors the beam power. A plot of this experimental arrangement is shown in Figure 2.

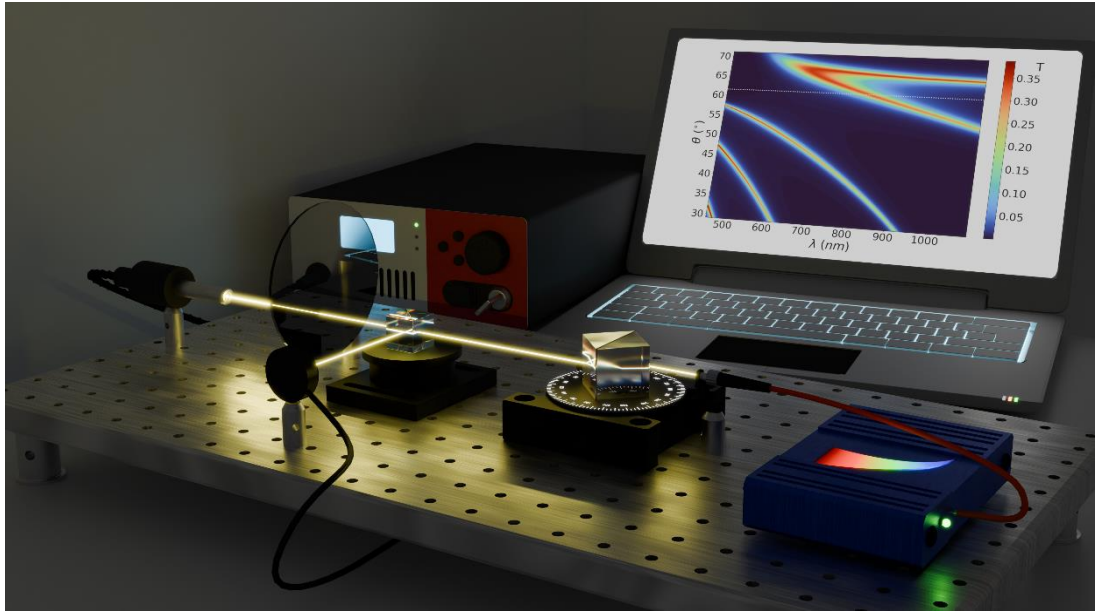

Figure 2. Experimental set-up. From the left to the right: Collimated Super-Continuum White Light Source (SCWLS), Variable Optical Density (VOD), Polarizing Beam Splitter (PBS), photodiode, motorized rotation platform, prism pair containing the MC (see fig. 1a), fibre spectrometer.

Measurements are made by rotating the platform while synchronously recording the spectrum of the light transmitted through the cavity. To calculate transmittance, we measure a reference spectrum of the light incident on the MC. In this way, we obtain a map of transmittances as a function of the angle of incidence, $\theta$, and wavelength, $\lambda$, of light, for a constant cavity thickness, $d$. Measurements were taken for two different intracavity fluids, directly with an air gap between prisms, and by infiltrating water. In Fig. 3 we show four examples of such measurements.

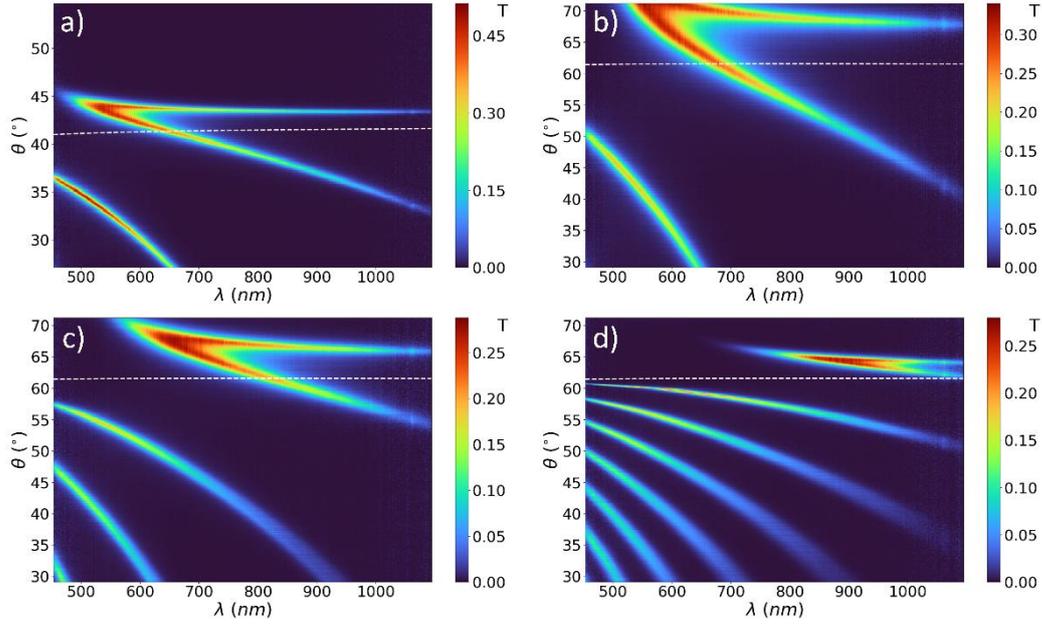

Figure 3. Transmittance ($T$) as a function of light wavelength ($\lambda$) and angle of incidence ($\theta$). (a) corresponds to an air gap of width $d \simeq 800\ nm$; (b), (c) and (d) correspond to $d \simeq 500, 800$ and $1500\ nm$ wide water gaps, respectively. In fact, these widths are obtained by comparison between experimental and theoretical maps (see below). The white dashed line corresponds to the critical angle curve of a HL interface, $\theta_{cr}$.

Fig. 3a shows the measured transmittance ($T$) with an air gap around $d \simeq 800\ nm$ between the mirrors. Three branches with high $T$ values can be observed, with their centres (maxima) defining the position of the cavity resonances. The two upper branches merge at a wavelength near 550 nm and at an angle of about 44°. The point at which these branches come together is called the point of coalescence, and for a given thickness it is determined by its coordinates $\theta_{co}$ and $\lambda_{co}$ (coalescence angle and wavelength respectively). The highest branch is always located above the critical angle, $\theta_{cr}$, corresponding to an HL interface (high index surrounding – low index intracavity media), which means that it is a CSP resonance. In fact, it is the fundamental or primary CSP resonance. All the resonances that spam under $\theta_{cr}$ correspond to FP resonances. An example is the lower branch in this plot. The third intermediate branch appears to be a FP resonance for long wavelengths.

However, it crosses the critical angle curve at a wavelength of 680 nm or so, transforming to a CSP type resonance. This means that it is a hybrid CSP/FP resonance: the primary FP resonance below $\theta_{cr}$, and the secondary CSP resonance above it. Another salient feature is that the higher CSP branch is mainly horizontal (fairly at the same incident angle) for wavelengths longer than that corresponding to coalescence. This means that there is a very broad CSP spectral resonance

Figs. 3b, 3c, and 3d display transmittance maps for water gaps of different thicknesses within the same $\theta - \lambda$ range. The map in Fig.3b, where the gap is around $d \approx 500$ nm, shows similar features to Fig.3a, but with the high transmittance branches located at larger incidence angles, due to water having a higher refractive index than air. The water gap in Fig. 3c is thicker, resulting in a displacement of the resonance branches towards longer wavelengths and in the appearance of two new FP resonances. Finally, Figure 3d shows the thickest water gap, so many more FP resonances are visible. Furthermore, the coalescence of the upper branches is further displaced to longer wavelengths. Comparing Figs. 3b – 3d, it is observed that the coalescence point varies with the MC thickness, ultimately describing a curve in the 3D space $(d, \theta, \lambda)$. Note also that the first pure FP resonance seems to run parallel to the critical angle curve at lower wavelengths. A detailed and analytical explanation of the behaviour of the resonances discussed so far is given below.

**3. Transmission theory**

3.1. Resonances

In order to calculate the transmission properties of the MC analytically, we consider a structure consisting of three parallel layers separated by planar interfaces. This stratified system can be mathematically modelled by iteratively considering the Fresnel coefficients at each of the interfaces and the propagation of the fields through each layer [23].

Restricting ourselves to the fully symmetric case (see Fig. 1a), we obtain the amplitude reflection and transmission coefficients of the whole system as follows:

$$r = r_{HML} + \frac{t_{HML}\, t_{LMH}\, r_{LMH}\, e^{i2k_{Ln}d}}{1 - r_{LMH}^2\, e^{i2k_{Ln}d}}$$
$$t = \frac{t_{HML}\, t_{LMH}\, e^{ik_{Ln}d}}{1 - r_{LMH}^2\, e^{i2k_{Ln}d}} \quad (1)$$

In this formula, $d$ is the intracavity thickness, and the amplitude reflection (transmission) coefficient $r_{ijk}$ ($t_{ijk}$) refers to a wave going from medium $i$ to medium $k$ by crossing medium $j$ (always a metal). These coefficients define the optical properties of the metallic mirrors. $k_{Ln}$ is the normal component of the wave vector in the intracavity medium. It is a function of $\theta$, the angle of incidence on the first metallic layer, i.e.:

$$k_{Ln} = k_0 \sqrt{n_L^2 - n_H^2 \sin^2 \theta}, \quad (2)$$

which is real below the critical angle corresponding to an HL interface, i.e., $\theta_{cr} = \sin^{-1}(n_L/n_H)$, and imaginary above it.

Our modelling of the MC is built around the transmittance, which for equal input and output media verifies $T = |t|^2$. The formula is well known when considering small angles of incidence [1], but it can be adapted for angles larger than the critical one:

$$T = \frac{|t_{LMH}|^2\, |t_{HML}|^2}{4\, |r_{LMH}|^2\, [\, \sinh^2(k_{Ln}''d - \ln|r_{LMH}|) + \sin^2(k_{Ln}'d + \varphi_{LMH})\, ]}, \quad (3)$$

where $\varphi_{LMH}$ is the phase of $r_{LMH}$, and we have decomposed the normal component of the wave vector into its real and imaginary parts, $k_{Ln} = k_{Ln}' + ik_{Ln}''$. This decomposition is relevant since only one out of the two is non-null for each angle of incidence ($k_{Ln} = k_{Ln}'$ for $\theta \leq \theta_{cr}$ and $k_{Ln} = ik_{Ln}''$ for $\theta > \theta_{cr}$).

Fig. 4 shows the transmittance maps in the $\theta - \lambda$ plane calculated by means of eq. (3), for the experimental conditions in Fig. 3. The RefractiveIndex.INFO database [24] was consulted for the optical constants of the materials used (silver, air, water and HK9). Specifically, the following references were used: [25] for silver, [26] air, [27] for water, [24] HK9. Theoretical maps were calculated for several intracavity thicknesses (with a resolution of 10 nm). Two-dimensional correlations were computed between each of theoretical map and the corresponding experimental version, selecting the thickness with the highest correlation as the most likely value. The similarity between experimental and theoretical graphs is evident, confirming the suitability of eq. (3) for modelling the transmittance of the MC. However, there are some differences. The experimental transmittances are generally smaller, especially for longer wavelengths. It is also noticeable that the experimental branches are wider, especially for smaller incident angles. Nevertheless, the positions of the theoretical resonances correctly reproduce the experimental ones.

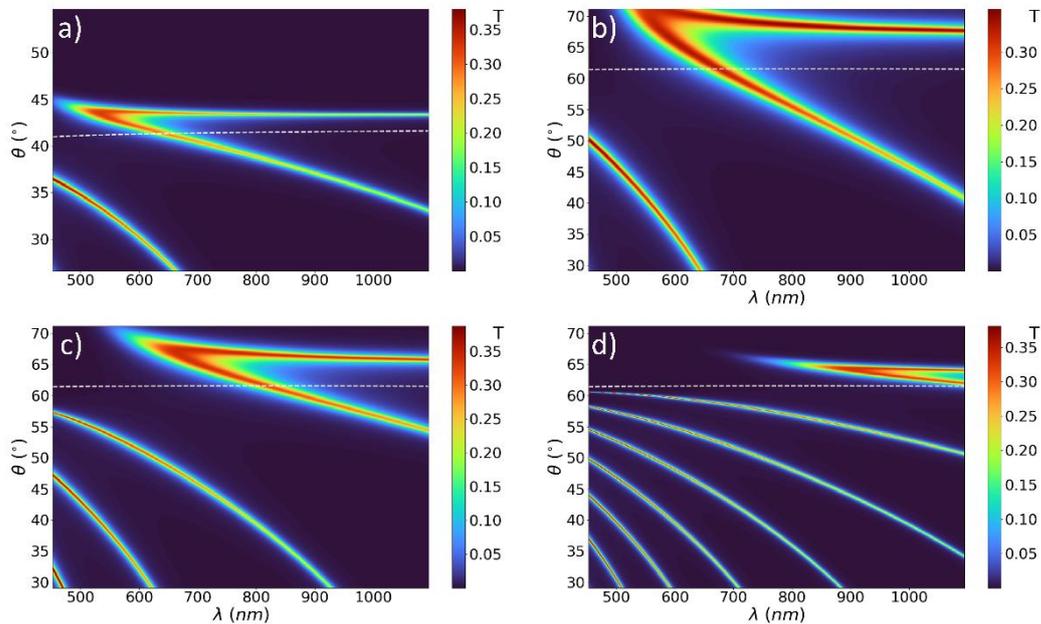

Figure 4. Theoretical transmittances corresponding to the experimental cases shown in Fig. 3. The white dashed line corresponds to the critical angle curve of a HL interface, $\theta_{cr}$.

Based on the transmission formula, we can derive the resonance conditions, corresponding to the transmission maxima. Considering that the characteristics of the metallic mirrors are specified, the transmittance depends on three variables: the wavelength of the light, $\lambda$, the angle of incidence in the cavity, $\theta$, and the thickness of the intracavity medium, $d$. For the sake of simplicity, in the following we assume a constant $\lambda$.

Below the critical angle, $k_{Ln} = k'_{Ln}$ is real and the only factor in eq. (3) that depends on the intracavity thickness is the sine argument. For a constant angle of incidence, the transmittance is maximum when the intracavity thickness $d$ cancels the sine function. The sine function equals zero whenever its argument is an integer multiple of $\pi$, so the resonant condition for maximum $T$ is:

$$k'_{Ln}d + \varphi_{LMH} = m\pi \quad m \in \mathbb{Z}^+, \quad \theta \leq \theta_{Hcr} \tag{4}$$

This equation is also approximately valid when the thickness is constant, and the angle of incidence is variable. Eq. (4) is the well-known phase condition for FP resonances [1]. It means that the phase accumulated by the wave in a complete oscillation in the cavity is a multiple of $2\pi$ at resonance.

On the other hand, above the critical angle $k_{Ln} = ik''_{Ln}$ is imaginary, and in this case the only factor that depends on the thickness is the argument of the hyperbolic sine. Therefore, at a constant angle of incidence, transmittance maxima are obtained when this function cancels out and the resonance condition is found to be:

$$k''_{Ln}d - \ln|r_{LMH}| = 0 \quad \Rightarrow \quad |r_{LMH}| = e^{k''_{Ln}d} \quad \theta \geq \theta_{Hcr} \tag{5}$$

Eq. (6) has a unique solution for a fixed angle of incidence, but the resulting value of $d$ can be repeated for two different angles of incidence, as shown in Fig. 5a. The smallest of these two angles of incidence corresponds to the secondary CSP resonance, while the largest angle corresponds to the primary CSP resonance. These resonances appear at $\theta = 0°$ and $\theta = 90°$, respectively, for different values of $d$ (smaller for the primary resonance), and gradually tend towards the value $\theta_{co}$, at which they converge or coalesce to a single resonance. This occurs for $d = d_{co}$, that can be calculated to be approximately:

$$d_{co} = max[\ln|r_{LMH}|/k''_{Ln}] \tag{6}$$

This equation also determines the coalescence angle $\theta_{co}$, which maximises $\ln|r_{LMH}|/k''_{Ln}$ at constant wavelength. Thereafter, for $d > d_{co}$ we have a single resonance, or better, two superposed degenerate resonances, with transmittance maxima that decrease as d increases. They do not correspond to solutions of eq. (5), but to a horizontal line with constant incident angle $\theta = \theta_{co}$.

There is a fundamental difference between FP resonances and CSP resonances, apart from verifying different conditions. FP resonances are bulk resonances corresponding to waves spread throughout the cavity. CSP resonances are surface resonances, where the EM fields are concentrated at the intracavity metal-dielectric interfaces but connected at the dielectric medium.

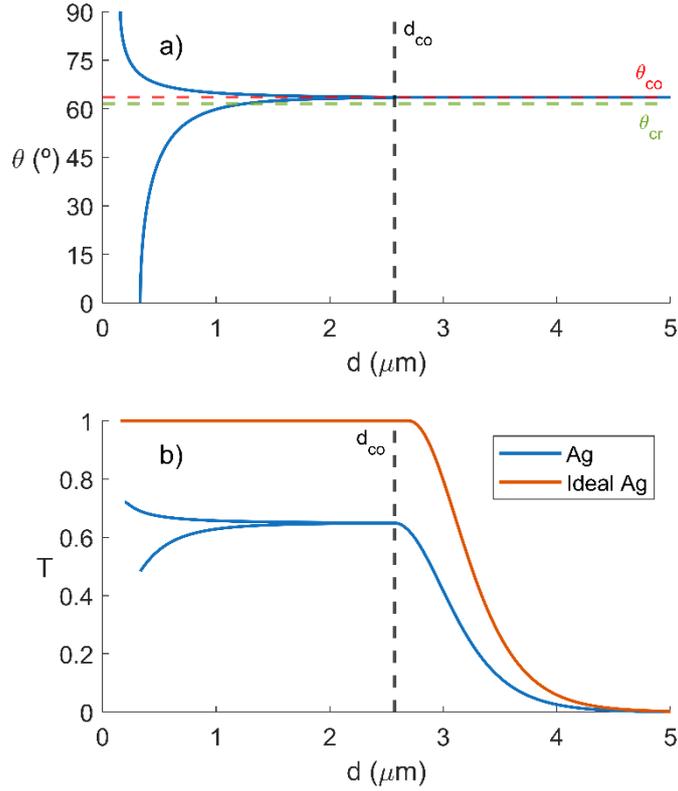

Figure 5. (a) Angle of incidence that maximises transmittance of the two CSP resonances as a function of MC intracavity thickness; (b) Values of transmittance at resonance as a function of MC thickness. For comparison, we also show the corresponding transmittance for a material that has a real electrical permittivity and is equal to the real part of the electrical permittivity of silver. The intracavity media is water and light wavelength is $\lambda = 1~\mu m$.

In Fig. 5a we plot the locus of pure and hybrid CSP resonances (the transmission maxima), versus cavity thickness, for a water gap and 1 micron wavelength. They are obtained using the resonance condition (4) with $m = 1$ for $\theta \leq \theta_{cr}$, and (5) for $\theta \geq \theta_{cr}$. As mentioned above, for distances greater than the coalescence distance, the resonance curve is completed by the horizontal line $\theta = \theta_{co}$. We have also calculated transmission maxima numerically from eq. (3), holding a constant $d$ and varying the incidence angle $\theta$. The differences between the two calculations are minimal, with the largest, though still very small, occurring near the coalescence point $(d_{co}, \theta_{co})$. We also show in Fig. 5b how the transmittance values at the resonances vary as a function of $d$. The maximum transmittance of the primary CSP resonance decreases gradually with $d$ until the coalescence thickness, $d_{co}$, is reached. At the same time, the maximum transmittance of the hybrid resonance increases from a lower initial transmittance until it also reaches the coalescence value. From this point on, the transmittances coincide and decrease gradually to zero. It is noteworthy that for thicknesses $d < 3.5$ µm (and $\lambda = 1$ µm), the transmittance is greater than 10%, i.e. it exceeds by several units the depth of penetration of light into the cavity (defined at the limit $d \to \infty$, for the same angle of incidence). As a comparison, we show the transmittance of a material with the same real part of the permittivity but with null imaginary part (it could be named "ideal silver" $\varepsilon_{Ag} = \varepsilon'_{Ag} < 0$). For the ideal metal the transmittance of the two CSP resonances is equal to unity up to the coalescence point (which occurs at a slightly greater thickness) and then decreases similarly to silver. While ideal, these are the best conditions for any application, since they correspond to the maximum possible transmittance.

Moreover, we can distinguish three regions in Fig. 5b. The first one, from $d = 0$ to $d \simeq d_{co}$, corresponds to a zone of high transmittance. This can be referred to as the cavity zone, where the only process that decreases the transmittance from its possible maximum value $T = 1$ is the absorption in the metal mirrors. The second region corresponds to thickness values greater than $d_{co}$, up to those values where the transmittance becomes extremely low, on the order of 1%. Within this range, the transmittance drops abruptly, mainly due to the frustrated total reflection process (in addition to absorption, of course). Thus, we call this the FTR zone. Finally, in the third region, the ATR zone, the almost irrelevant transmittances decrease gradually and slowly towards zero, so the main process is reflection attenuated by absorption.

3.2 Resonance areas

The discussion can be continued by referring to Figure 6. In this figure, two dispersion curves -corresponding to the critical and coalescence angles- divide the $\lambda - \theta$ plane into three different areas. The area below the critical angle curve (I) corresponds to the FP domain, while all the area above this curve may be referred to as the SP domain. Within this SP domain there are two different area s, delimitated by the coalescence angle curve: the area between the two curves (II) is the locus of the secondary CSP resonance, while the top area (III) corresponds to the locus of the primary CSP resonance. The graph could be completed by displaying different resonance curves. For example, those corresponding to Figures 2 or 3, i.e. the maximum transmission curves for the different branches. As another example, it would also allow us to map the evolution of a resonance as a result of varying certain parameters, such as the intracavity thickness or the refractive index.

The previous discussion for constant wavelengths corresponds to vertical lines in Fig. 6. Thus, a FP resonance begins at normal incidence for a given $d$ and gradually shifts towards the critical angle as $d$ increases. Meanwhile the hybrid resonance crosses the critical angle curve towards the coalescence angle curve, which is reached at $d = d_{co}$. In contrast, the primary CSP resonance starts at $\theta = 90°$ for small $d$, and ends at $\theta_{co}$ for larger thicknesses. Alternatively, if we consider a fixed incident angle, we can follow the resonances in the spectral domain as $d$ varies. The corresponding resonance curves are horizontal cuts in Fig. 6. In this case, FP resonances only occur at each wavelength for fixed angles below the critical angle curve, shifting from short to longer wavelengths as $d$ increases. For higher angles, the two possible resonances are always CSPs. Both can be seen to converge to the coalescence angle curve from the two sides of the spectrum when $d = d_{co}$. As shown in the inset, for constant $\theta$ we can only talk about hybrid resonance if $\theta$ equals the critical angle for some wavelength in our range, cutting the critical angle curve. In this case, the resonant wavelength increases as $d$ does so.

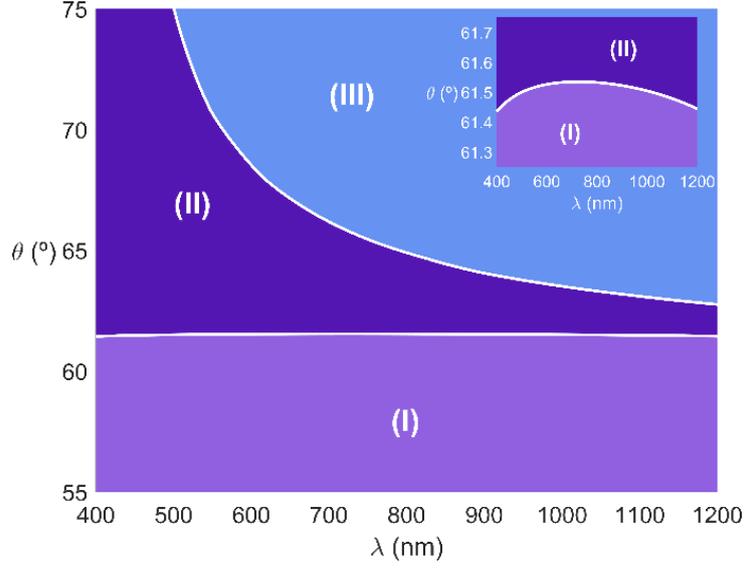

Figure 6. Map of resonances separated in three areas by the curves of angles $\theta_{cr}$ and $\theta_{co}$ as a function of wavelength. The surrounding dielectric $n_H$ is HK9, and the intracavity medium is water.

3.3 The amplitude reflection coefficient

Returning to the expression of transmittance, eq. (3), it is clear that the amplitude coefficient $r_{LMH}$ is the key parameter determining the MC resonant behaviour, especially in the SP regime. This coefficient depends on the Fresnel coefficients at the metal-dielectric boundary as:

$$r_{LMH} = \frac{-r_{ML} + r_{MH}\, e^{i\phi_M}}{1 - r_{ML}\, r_{MH}\, e^{i\phi_M}}$$
$$\phi_M = 2d_M \cdot k_{Mn} = 2d_M \cdot k_0 \sqrt{\varepsilon_M - n_H^2 \sin^2 \theta_H} \qquad (7)$$

where $d_M$, $\varepsilon_M$ are metal thickness and complex permittivity, respectively. $k_{Mn}$ corresponds to the component of the wavevector in the metal normal to the interfaces. In Fig. 7 we plot the absolute value and phase of $r_{LMH}$, both as a function of angle of incidence and wavelength. For comparison, the absolute value and phase of $r_{LM}$ are also shown.

$r_{LM}$ corresponds to the limit of $r_{LMH}$ as $d_M \to \infty$. It is also the key parameter for the generation of simple SPR in a single metal-dielectric interface [28].

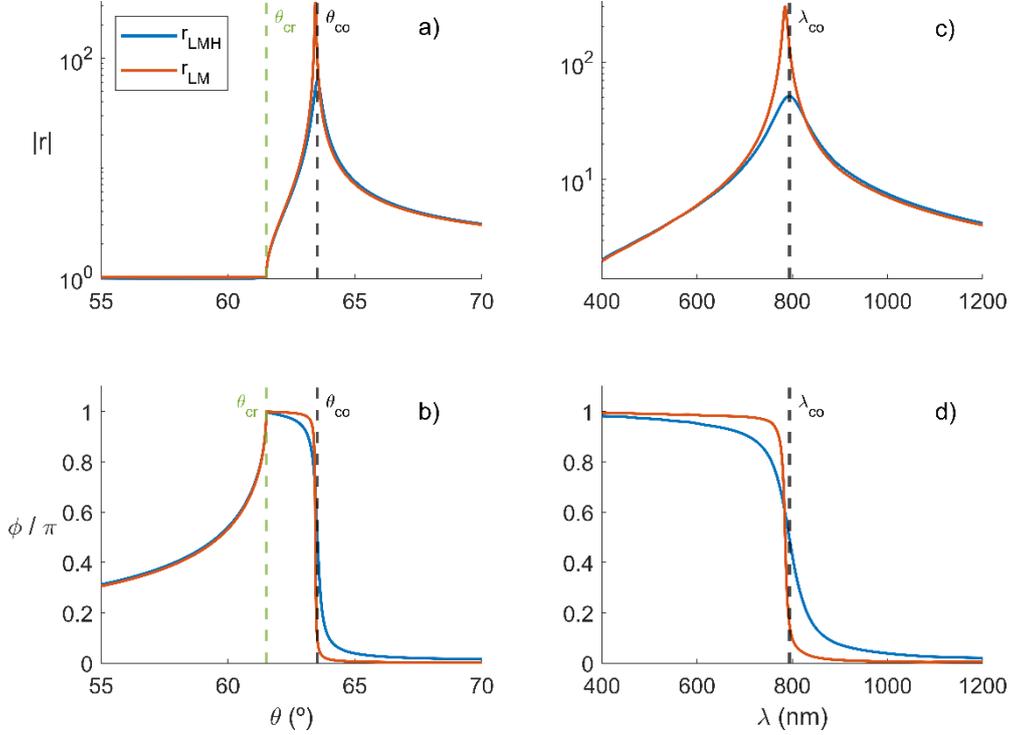

Figure 7. Absolute value and phase of $r_{LMH}$ (blue) and $r_{LM}$ (red) as a function of incident angle and wavelength for $d_M = 45$ nm. $\lambda = 1$ μm in (a, b) and $\theta = 65°$ in (c, d). We have considered again water as the intracavity fluid between two silver layers, with HK9 glass as the surrounding $n_H$ medium.

Considering Fig. 7a, it is evident that $|r_{LMH}|$ behaves very differently below and above the critical angle. Below $\theta_{cr}$, $|r_{LMH}|$ stays below and very close to 1, while above $\theta_{cr}$, it has a value much greater than 1, with a maximum at the coalescence angle. The shape of $|r_{LMH}|$ varies with the thickness of the metallic mirrors, converging to $|r_{LM}|$ for large thicknesses. The behaviour of $|r_{LM}|$ has been studied in ref. [28]. At first sight, the fact that $|r_{LM}|$ and $|r_{LMH}|$ are greater than one may seem surprising, since it would seem to break the conservation of energy. This is not the case, as shown for example in ref. [29]. Concerning the phase, in Fig. 7b, it grows continuously for small angles until reaching $\pi$ at the critical angle. Thereafter, it slowly decreases until near the angle of coalescence it drops abruptly to a value close to zero, and then decreases slowly again. Furthermore, exactly at $\theta_{co}$ the phase is $\pi/2$. In the case of $r_{LM}$, the phase behaviour is very similar below the critical angle and has a steeper drop around the coalescence value. On the other side, the wavelength dependence shown in Fig. 7c and 7d is comparable to that discussed for angles greater than the critical angle in Fig. 7a and 7b. This happens because the plots on the right correspond to a fixed angle above the critical angle curve in the spectral range plotted, ($\theta_{cr} < 61.6°$ for any wavelength within that range, for example $\theta_{cr} = 61.55^o$ for $\lambda = 1$ μm). If we had chosen a constant angle below, the module of the reflection coefficients would be lower than 1, and the variations in their phase would be smoother. Finally, note that $\theta_{co}$ and $\lambda_{co}$ vary slightly with metal thickness.

## 4. Conclusions

To summarize, we have examined the transmission of an FP type microcavity with metal mirrors for any angle of incidence and optical wavelengths in the uncommon configuration of a surrounding medium with greater refractive index than the intracavity medium. We concentrate our analysis on the SP regime where unusual transmission can be observed at resonance where CSP are excited, even if the cavity is several wavelengths thick. A hybrid resonance has been identified, which is an FP type resonance for incident angles under $\theta_{cr}$ and an SP type resonance for larger angles. Finally, the amplitude reflection coefficient $r_{LMH}$ is recognized as the key parameter determining the CSP resonance behaviour.

We hope that the analytical model detailed in this paper can be complemented by further studies addressing questions such as the energy flow in MC (including reflectance); how small changes in refractive index -and also absorption- affect the reflectivity and transmissivity of MC; or the design of complex mirror structures to improve the optical properties of MC. We also hope that this study will pave the way for investigating new/better applications of such devices in fields such as optical sensing, spectral filtering or surface and film analysis.

**Acknowledgments:**

Y. Arosa acknowledges the funding from the postdoctoral fellowship ED481D-2024-001 from Xunta de Galicia. A. Doval would like to thank Ministerio de Universidades for the financial support through FPU21/01302. This research is funded by the contract USC 2024-PU031 from the University of Santiago de Compostela and GRC ED431C 2024/06 from Xunta de Galicia.

**Conflict of interest**

The authors have no conflict of interest related to this publication.